\documentclass[pra,twocolumn,amsmath,showpacs]{revtex4}
\usepackage{graphicx}
\usepackage{bm}
\begin{document}

\title{Heteronuclear fermionic superfluids with spin degrees of freedom}
\author{D. B. M. Dickerscheid$^{1}$}
\email{dd1978@gmail.com}
\author{Y. Kawaguchi$^{1}$}
\author{M. Ueda$^{1,2}$}
\affiliation{$^1$Department of Physics, Tokyo Institute of Technology,
2-12-1 Ookayama, Meguro-ku, Tokyo 152-8551, Japan}
\affiliation{$^2$ERATO, Macroscopic Quantum Control Project, JST, Bunkyo-ku,
Tokyo 113-8656, Japan}
\date{\today}
\begin{abstract}
We present a theory of spinor  superfluidity
in a two-species heteronuclear ultracold fermionic atomic gas 
consisting of arbitrary half-integer spin and one-half
spin  atoms. 
In particular, we focus on the magnetism of the superfluid phase
and determine the possible phases in the absence of 
a magnetic field.
Our work demonstrates similarities between 
heteronuclear fermionic superfluids and
 spinor Bose-Einstein condensates at the mean-field level.
Possible experimental situations are discussed.
 \end{abstract}
\maketitle

\section{Introduction.}

Fermionic superfluidity is currently one of the 
most active research topics in the field of ultracold atomic gases.
Fermionic superfluids have been realized and 
the BEC-BCS crossover regime has been explored 
for both balanced and imbalanced systems
~\cite{BCS1,BCS2,Jochim,Regal,BCS3,BCS4,partridge1,zwierlein1,zwierlein2}.
Another line of research that has attracted a growing interest
concerns mixtures of different atomic species 
\cite{Truscott,Schreck,Hadzibabic,Roati,Ospelkaus}
such as two-species
Fermi-Fermi mixtures.
Fermionic heteronuclear gases have currently 
been investigated both
experimentally  \cite{Ewille} 
and theoretically \cite{Bedaque,Liu,iskin:100404,Lin}.
Recently, the first quantum degenerate 
two-species Fermi-Fermi mixture 
has been realized \cite{Dieckmann1}.

In this paper we theoretically 
investigate 
superfluid properties of a heteronuclear Fermi-Fermi 
mixture with spin degrees of freedom.
Two-species Fermi mixtures 
differ from the single-species gas in such a fundamental way that
at low temperatures
atoms in the same spin state can
 interact with each other through $s$-wave scattering. 
For identical fermions, however,
this is prohibited due to the Pauli principle and 
consequently superfluidity of a 
single species of atoms occurs in the spin-singlet state for the 
$s$-wave channel.
Specifically, for the case
where both atoms have spin one-half this means that 
both spin-singlet and spin-triplet states are allowed.
Theoretical investigations of superfluidity of 
heteronuclear Fermi gases have so far not explored the physics 
of the  (hyperfine) spin degrees of freedom of 
the atoms \cite{iskin:100404}.
While pairing in single-species
fermionic superfluids with 
arbitrary spin has been studied \cite{JasonHo1, Honerkamp},
no attempt to investigate the combined effects of those two, i.e.,
 heteronuclear spinor superfluidity, has been made to the best of our knowledge.
This is the subject we address in this work. In particular,
our work demonstrates a strong similarity between heteronuclear superfluids
and spinor BECs at the mean-field 
level and suggests the existence of novel many-body states.

Experimentally, a particularly interesting candidate to
realize a heteronuclear spinor superfluid is an isotope
 mixture of the rare-earth element ytterbium (Yb),
which has two stable fermionic isotopes
with nuclear spin $I=1/2$ (\mbox{$^{171}$Yb}) and 
$I=5/2$ (\mbox{$^{173}$Yb}), respectively, 
and electronic 
spin $S=0$. Both isotopes have already 
been trapped optically and recently the $^{173}$Yb gas was cooled 
to quantum degeneracy \cite{Takahashi1}. 

This paper is organized as follows.
We formulate our problem in Sec. \ref{sectheory} 
and develop the Bogoliubov and Ginzburg-Landau theories for
our system in  Secs. \ref{secthbog} and \ref{secglth}, respectively.
We apply the theory for specific cases in Sec. \ref{secres}
and discuss possible experimental realization of 
the gases considered here in Sec. \ref{secexp} .
 We summarize and conclude this 
paper in Sec. \ref{secsum}. Some algebraic manipulations to derive the 
Bogoliubov excitation spectrum of a heteronuclear superfluid have been
relegated to the Appendix to avoid disgressing from the main subject.

\section{Theory of Heteronuclear Fermionic Superfluidity}

\subsection{Formulation of the problem}\label{sectheory}
We consider an optically trapped Fermi-Fermi mixture with spins
$f_{\phi} \geq 1/2$ and $f_{\chi} = 1/2$, where
we distinguish the two atomic species with suffixes 
$\phi$ and $\chi$.
 We assume that the temperature is sufficiently low so that 
we have only to consider $s$-wave interactions between the atoms.
In general, there are interspecies and intraspecies interactions, and
experimentally it is possible to 
tune the interspecies interactions independently of the  
intraspecies interactions. In this paper we will ignore
the intraspecies interactions to focus on the bare essentials
of this system.
The total spin $f$ of the interacting atoms with spin 
 $f_{\phi} \geq 1/2$ and  $f_{\chi} = 1/2$ is given by 
$f^{\pm} \equiv f_{\phi} \pm 1/2$.
In the absence of an external magnetic field,
the total Hamiltonian $\hat{H} = \hat{H}_{1B} + \hat{V}$
is the sum of one-body part
$\hat{H}_{1B}$
and interaction $\hat{V}$.
The general form of the interaction can be represented as
\begin{eqnarray}\label{genpot}
\hat{V} = \sum_{f=f^{\pm}} {V}_{f} (\bm{x} - \bm{x}')\mathcal{\hat{P}}_{f},
\end{eqnarray}
where the projection operators 
$\mathcal{\hat{P}}_{f}$ project  pairs of atoms onto pairs of
the total spin 
$f$ channel and ${V}_{f} ({\bm x} - {\bm x}')$ 
is the interaction potential between a pair of atoms with total spin 
$f$. For $s$-wave interactions,
 the interaction potential can be approximated by a pseudopotential
$V_{f}(\bm{x} - \bm{x}') = g_{f} \delta(\bm{x} - \bm{x}')$.

In second quantization 
the one-body Hamiltonian is given by
\begin{eqnarray}\label{eq1}
\hat{H}_{1B} &=& \! \! \int \! d{\bm{ x}} \sum_{\sigma = -f_{\phi}}^{f_{\phi}}
\hat{\phi}^{\dagger}_{\sigma} ({\bm x})
 \left(
- \frac{\hbar^{2} 
%\boldsymbol
{\nabla}^{2}}{2 M_{\phi}} 
+ V_{\phi,\sigma}^{\rm ex} ({\bf x}) - \mu_{\phi}
\right) 
%\delta_{m,n} +
%p
%{^{j}}F^{{z}}_{m,n}
\hat{\phi}^{\phantom \dagger}_{\sigma} ({\bm x})
\nonumber \\ 
&+&
\! \! \int \! d\bm{x} \sum_{\sigma=\pm 1/2}
\hat{\chi}^{\dagger}_{\sigma} (\bm{x})
\left(
- \frac{\hbar^{2} 
%\boldsymbol
{\nabla}^{2}}{2 M_{\chi}} 
+ V_{\chi, \sigma}^{\rm ex} (\bm{x}) - \mu_{\chi}
\right) 
%\delta_{\sigma,\tau} +
%p
%{^{j'}}F^{{z}}_{\sigma,\tau}
\hat{\chi}^{\phantom \dagger}_{\sigma} (\bm{x}), \nonumber \\ 
\end{eqnarray}
where $\hat{\phi}^{\dagger}_{\sigma}$ ($\hat{\phi}_{\sigma}$) and 
 $\hat{\chi}^{\dagger}_{\sigma}$ ($\hat{\chi}_{\sigma}$) are 
 the creation (annihilation) operators of
atoms of spin $f_{\phi}$ and spin $1/2$, respectively,
 in the magnetic sublevel $\sigma$.

The terms $ V_{\phi,\sigma}^{\rm ex} (\bm{x})$ and
$V_{\chi,\sigma} ^{\rm ex} (\bm{x})$
describe state-dependent external potentials for the $\phi$ and 
$\chi$ atoms, respectively, $\mu_{\phi}$ and $\mu_{\chi}$
are the chemical potentials,
and $M_{\phi}$ and $M_{\chi}$ are the  masses of the
$\phi$ and $\chi$ atoms, respectively.
The projection operators can be expressed in second 
quantisation as 
\begin{eqnarray}
\mathcal{\hat{P}}_{f} = \sum_{m}
 \hat{A}^{\dagger}_{f,m} (\bm{x}) \hat{A}_{f,m} (\bm{x}),
\end{eqnarray} 
where
\begin{eqnarray*}
 \hat{A}_{f,m} (\bm{x}) = 
\sum_{\sigma,\sigma'} 
\langle f,m | f_{\phi}, \sigma; 1/2, \sigma' \rangle
\hat{\phi}_{\sigma} (\bm{x})\hat{\chi}_{\sigma'} (\bm{x})
\end{eqnarray*}
is the annihilation operator 
of a pair of $\phi$ and $\chi$ atoms with total spin $f$ and 
total magnetic quantum number $m$,
with
$\langle f,m | f_{\phi}, \sigma; 1/2, \sigma' \rangle$ being
a Clebsch-Gordan coefficient.
The interaction Hamiltonian is then given by
\begin{eqnarray}
\hat{V} =
\frac{1}{2}  \sum_{f,m} \int d\bm{x}~ 
 g_{f}  \hat{A}^{\dagger}_{f,m} (\bm{x})  \hat{A}^{\phantom \dagger}_{f,m} (\bm{x}).
\end{eqnarray}
Since we are interested in the superfluid phase we assume
the interaction between the atoms to be attractive, i.e.,
$g_{f}<0$.

The interaction Hamiltonian
 in Eq. (\ref{genpot}) can be rewritten in a physically suggestive manner.
It follows from the completeness relation
 $1 =  \mathcal{\hat{P}}_{f^{+}}
+  \mathcal{\hat{P}}_{f^{-}}$ and from the total spin squared
\begin{eqnarray*}
( {\bm f}_{\chi} + {\bm f}_{\phi})^{2} = 
\sum_{f=f^{+},f^{-}} f (f+1) \mathcal{\hat{P}}_{f},
\end{eqnarray*}
 with
$f^{\pm} \equiv f_{\phi} \pm 1/2$ that
\begin{eqnarray}
 \mathcal{\hat{P}}_{f^{\pm}} = \frac{1}{2 f^{+}}
\left[
f^{+} \pm \frac{1}{2} \pm 2  {\bm f}_{\chi} \cdot {\bm f}_{\phi}
\right].
\end{eqnarray}
The interaction Hamiltonian can then be constructed as
\begin{eqnarray}
\hat{V} &=& 
\left(
g_{f^{+}} \mathcal{\hat{P}}_{f^{+}} + g_{f^{-}} \mathcal{\hat{P}}_{f^{-}}
\right) \delta ({\bm x} - {\bm x}')~
\nonumber \\ &=&
 \left\{
g_{f^{+}} \frac{f^{+} + \frac{1}{2}}{2 f^{+}} 
+
g_{f^{-}} \frac{f^{+} - \frac{1}{2}}{2 f^{+}} 
\right\} \delta ({\bm x} - {\bm x}') \nonumber \\ &&+
\frac{(g_{f^{+}} - g_{f^{-}})}{f^{+}} \delta ({\bm x} - {\bm x}')~
 {{\bm f}}_{\chi} \cdot  {{\bm f}_{\phi}}.
\end{eqnarray}
The corresponding second-quantized expression is 
\begin{eqnarray}
&\hat{V}& =
%\frac{1}{2} \int d\bm{x} \sum_{f=f^{\pm}} g_{f} \left\{
%   \frac{:{^{1/2}} {\bf F} \cdot {^{j}} {\bf F}:}{f^{+}}  + 
%\left( \frac{1}{2} \pm \frac{1}{4 f^{+}} \right) : \hat{n}_{\chi} (\bm{x})\hat{n}_{\phi} (\bm{x}) : \right\}
%\nonumber \\
%&=&
\frac{1}{2} \int d\bm{x} 
\frac{(g_{f^{+}} - g_{f^{-}})}{f^{+}}
  :{^{\chi}} {\hat{\bm F}} ({\bm x})\cdot {^{\phi}} {\hat{\bm F}} ({\bm x}):
\nonumber \\
&+& 
\frac{1}{2} \int d\bm{x} 
\left\{
g_{f^{+}} \frac{f^{+} + \frac{1}{2}}{2 f^{+}} 
+
g_{f^{-}} \frac{f^{+} - \frac{1}{2}}{2 f^{+}} 
\right\}
: \hat{n}_{\chi} (\bm{x})\hat{n}_{\phi} (\bm{x}) :,
\nonumber \\
\end{eqnarray}
where $::$ denotes  normal ordering,
$\hat{n}_{\phi} (\bm{x})$ and 
$\hat{n}_{\chi} (\bm{x})$ are the
total densities of the $\phi$ and $\chi$ atoms given by
\begin{eqnarray}
\hat{n}_{\phi} (\bm{x}) = 
\sum_{\sigma} \hat{\phi}^{\dagger}_{\sigma} (\bm{x}) 
 \hat{\phi}^{\phantom \dagger}_{\sigma} (\bm{x}),
\end{eqnarray}
and
\begin{eqnarray}
\hat{n}_{\chi} (\bm{x}) = 
\sum_{\sigma} \hat{\chi}^{\dagger}_{\sigma} (\bm{x}) 
\hat{\chi}^{\phantom \dagger}_{\sigma} (\bm{x}) 
,
\end{eqnarray} and
${^{\chi}} {\hat{\bm F}} ({\bm x})$
and ${^{\phi}} {\hat{\bm F}} ({\bm x})$ denote
 the  spin density vectors whose components are given by
\begin{eqnarray}
^{\phi} \hat{ F}^{i} (\bm{x}) = 
 \sum_{\sigma,\sigma'}
\hat{\phi}^{\dagger}_{\sigma}(\bm{x})
 \left[ {^{\phi}}F^{i }\right]_{\sigma \sigma'}
\hat{\phi}_{\sigma'} (\bm{x})
\end{eqnarray}
 and
\begin{eqnarray}
^{\chi} \hat{ F}^{i} (\bm{x}) =
 \sum_{\sigma, \sigma'}
\hat{\chi}^{\dagger}_{\sigma} (\bm{x})
\left[ {^{\chi}}F^{i} \right]_{\sigma \sigma'} 
\hat{\chi}_{\sigma'} (\bm{x})
,\end{eqnarray} 
respectively, 
where 
$[{^{\phi}}{F}^{i}]_{\sigma \sigma'}$
 and $[{^{\chi}}{F}^{i}]_{\sigma \sigma'}$
are the matrix elements 
of the $i=x,y,z$ components of  
spin matrix vectors
$^{\phi}{\bm F}$ and
$^{\chi}{\bm F}$.
The sign of $(g_{f^{+}} - g_{f^{-}})$ specifies which
of the two spin states $f^{+}$ and $f^{-}$
is energetically favorable and thus
determines if the total angular momentum
of the Cooper pairs is $f^{+}$ or $f^{-}$.
\begin{figure}
\includegraphics[width=.85\columnwidth]{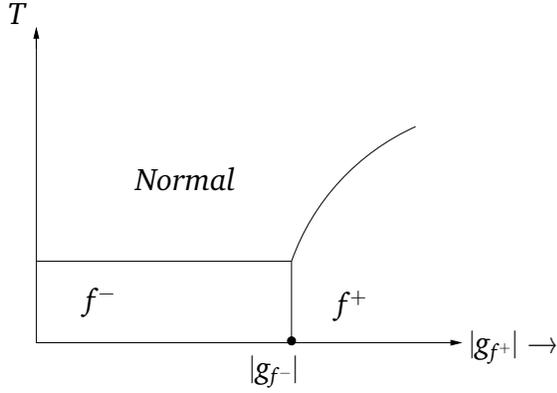}
\caption{Generic phase diagram as a function of
temperature $T$ and coupling constant $g_{f^{+}}$ . 
 $f^{-}$ ( $f^{+}$) denotes the phase in which 
 the spin angular momentum of a constituent Cooper pair is $f^{-}$
 ( $f^{+}$). 
}\label{fig1}
\end{figure}
The generic phase diagram that results is sketched 
in Fig. \ref{fig1}. At high temperatures, the system is 
in the normal state and the total angular momentum is not fixed.
When we lower the temperature for a fixed value of 
$g_{f^{-}}$ and  
$|g_{f^{+}}|> |g_{f^{-}}|$,
we expect a phase transition to a 
BCS superfluid whose  total angular momentum of each Cooper pair
is $f^{+}$ and the transition temperature is determined by $g_{f^{+}}$.
On the other hand, when $|g_{f^{+}}| < |g_{f^{-}}|$, we expect a
BCS superfluid with total angular momentum $f^{-}$.
By changing the ratio $g_{f^{+}}/g_{f^{-}}$ we can switch between
the two types of superfluidity.

\subsection{Mean-field theory of heteronuclear superfluidity}\label{secthbog}

The phases we found in the previous subsection possess
 the internal spin degrees of freedom.
To investigate the magnetic properties
of the superfluid phases 
we make use of mean-field theory.
The order parameter is defined by
\begin{eqnarray}\label{defdm}
\Delta_{f,m} (\bm{x}) 
\equiv g_{f}\langle \hat{A}_{f,m} ( \bm{x}) \rangle .
\end{eqnarray}
Applying the Gorkov decoupling, we decompose the interaction Hamiltonian as
\begin{eqnarray}\label{mfV}
\hat{V} = 
\frac{1}{2}
\int d\bm{x}  \sum_{m}
\left( \hat{A}^{\dagger}_{f,m} (\bm{x})
\Delta_{f,m} + \Delta_{f,m}^{*} \hat{A}_{f,m}
- \frac{|\Delta_{f,m}|^{2}}{g_{f}} \right). \nonumber \\
\end{eqnarray} 
In a homogeneous system, the resulting mean-field Hamiltonian 
 can be diagonalized by means
of a Bogoliubov transformation. 
We find that 
the Bogoliubov dispersion relations are  given by (see Appendix 
\ref{appBD} for derivations)
\begin{eqnarray}\label{bogdisp}
&& \hbar \omega^{\pm}_{\phi, \bf k} =  \frac{1}{2} \left( 
\xi_{\phi, {\bf k}} - \xi_{\chi, {\bf k}} \right) 
+ \Omega_{\pm}
\end{eqnarray}
and
\begin{eqnarray}\label{bogdisp2}
&& \hbar \omega^{\pm}_{\chi, \bf k} =  \frac{1}{2} \left( 
\xi_{\chi, {\bf k}} - \xi_{\phi, {\bf k}} \right) 
+ \Omega_{\pm},
\end{eqnarray}
where
 $\xi_{\alpha, {\bf k}} = \epsilon_{\alpha, {\bf k}} - \mu_{\alpha}$,
$\epsilon_{\alpha, {\bf k}} = \hbar^{2} k^{2} / 2 M_{\alpha}$
($\alpha= \phi,\chi$),
and
\begin{eqnarray}
\Omega_{\pm} = 
\frac{1}{2} \sqrt{
\frac{1}{2} |\Delta|^{2}
 \pm \frac{ |{\bf F}_{\Delta}|}{2 {f^{+}}}  +
 \left( \xi_{\phi, {\bf k}} + \xi_{\chi, {\bf k}} 
\right)^{2}
}. \nonumber \\
\end{eqnarray}
Here
$|\Delta|^{2} =\sum_{m} |\Delta_{f,m}|^{2}$, the spin vector
$ {\bm F}_{\Delta} $ 
 of the Cooper pairs is given by
\begin{eqnarray}\label{cpspin}
{\bm F}_{\Delta}
= {\bm \Delta}^{\dagger} \cdot
{\bm F} \cdot {\bf \Delta}
= \sum_{m,m'}
\Delta^{*}_{f,m} 
\left( {\bf F} \right)_{m,m'}
\Delta_{f,m'},  
\end{eqnarray}
with $\left( {\bf F} \right)_{m,m'}$
being the spin-$f$ matrix, and
$\Delta_{f,m}$ is given in Eq. (\ref{defdm}).

In the absence of the spin degrees of freedom
the Bogoliubov modes in Eqs. 
(\ref{bogdisp}) and (\ref{bogdisp2}) describe
quasi-particle excitations of fermions with mass and 
population imbalance \cite{Lin,Bedaque,Liu}.
To determine the spin structure we next 
derive and solve the gap equation for our system.
After the Bogoliubov transformation the Hamiltonian is diagonalized as
\begin{eqnarray}\label{bogham}
\hat{H} &=& 
\sum_{\sigma,{\bf k}} \hbar \omega_{\phi,\bf k}^{\sigma} 
\hat{\eta}^{\dagger}_{\sigma, {\bf k}} \hat{\eta}_{\sigma,{\bf k}}
+
\sum_{\sigma,{\bf k}} \hbar \omega_{\chi,\bf k}^{\sigma} 
\hat{\zeta}^{\dagger}_{\sigma, {\bf k}} \hat{\zeta}_{\sigma,{\bf k}}
\nonumber \\ &&
+ \sum_{\bf k}
\left(
2 \xi_{\chi,\bf k} - \hbar \omega^{+}_{\chi,\bf k} 
- \hbar \omega^{-}_{\chi, \bf k} \right)
- \sum_{m} \frac{|\Delta_{f,m}|^{2}}{2 g_{f}}, \nonumber \\
\end{eqnarray}
where $ \hat{\eta}_{\sigma,{\bf k}}$ and
$ \hat{\zeta}_{\sigma,{\bf k}}$ are the Bogoliubov-mode operators
that reduce continuously to 
$\hat{\phi}_{\sigma,{\bf k}}$ and $\hat{\chi}_{\sigma,{\bf k}}$,
respectively, as $|\Delta|\rightarrow 0$.
There are additional contributions to the above Hamiltonian
arising from the degenerate unpaired modes with ideal gas 
dispersion $\xi_{\phi,{\bf k}} =
\hbar^{2} k^{2}/2 M_{\phi} - \mu_{\phi}$; these terms, however,
do not depend on the energy gap and are therefore not relevant 
for calculating the gap equation.
The gap equation follows from the 
thermodynamic potential $\Omega$, which is determined from the 
partition function $Z = e^{-\beta \Omega} = {\rm Tr} e^{-\beta \hat{H}}$.
Using the Hamiltonian in Eq. (\ref{bogham}), we obtain
\begin{eqnarray}
&&\Omega = 
 \sum_{\bf k}
\left(
2 \xi_{\chi,\bf k} - \hbar \omega^{+}_{\chi,\bf k} 
- \hbar \omega^{-}_{\chi, \bf k} \right)
- \sum_{m} \frac{|\Delta_{f,m}|^{2}}{2 g_{f}}
\nonumber \\ &&
- \frac{1}{\beta} \sum_{{\bf k},\alpha}
\log{\left( 1 + e^{-\beta \hbar \omega^{+}_{\alpha, \bf k}} \right)}
- \frac{1}{\beta} \sum_{{\bf k},\alpha}
\log{\left( 1 + e^{-\beta \hbar \omega^{-}_{\alpha, \bf k}} \right)}
. \nonumber \\
\end{eqnarray} 
The gap equations that follow from
$d\Omega/d\Delta^{*}_{f,m} = 0$ are given by
\begin{eqnarray}\label{gapeqn}
&&-\frac{\Delta_{f,m}}{T^{2B}_{f}} =
\sum_{l={\pm 1}} \sum_{\bf k}
\left\{
 \frac{\Delta_{f,m}}{2} \frac{1}{\epsilon_{\phi,{\bf k}} + \epsilon_{\chi,{\bf k}}}
 \right.
\nonumber \\ && \left.  
+ \frac{
N( \hbar \omega^{l}_{\phi, \bf k}) + 
N( \hbar \omega^{l}_{\chi, \bf k}) - 1
}{
2 \left( \hbar \omega^{l}_{\phi,\bf k } 
+ \hbar \omega^{l}_{\chi,\bf k}
\right)
}
\left(
\Delta_{f,m} +
 \frac{l}{2 {f^{+}}} \frac{d|{\bf F}_{\Delta} |}{ d \Delta^{*}_{f,m} }
\right)
\right\}
. \nonumber \\
\end{eqnarray}
Here by way of renormalization we have replaced the
 bare coupling constants
$g_{f}$ with the two-body $T$-matrix
$T^{2B}_{f}$, where $T^{2B}_{f}$ is related to 
$g_{f}$ by
\begin{eqnarray}
\frac{1}{T^{2B}_{f}} = -\frac{1}{g_{f}} 
+ \sum_{\bf k} (\epsilon_{\phi,\bf k} + \epsilon_{\chi, \bf k})^{-1}.
\end{eqnarray}
From an experimental point of view,
the two-body $T$-matrix is directly related to 
the $s$-wave scattering length $a_{f}$
with the total angular momentum $f$ of the two
colliding atoms 
via
$T^{2B}_{f} = 2 \pi a_{f} \hbar^{2} /M_{\rm r}$,
where
$M_{\rm r} = M_{\phi} M_{\chi} / (M_{\phi} + M_{\chi})$ is the reduced mass. 
If $\Delta_{f,m} \neq 0$ we can divide both sides 
of the gap equation by $\Delta_{f,m}$ and, as a result,
the left-hand-side of the gap 
equation (\ref{gapeqn}) becomes independent of $m$.
To be consistent, the right-hand-side should also be independent of 
$m$, so 
$
\frac{1}{\Delta_{f,m}} \frac{d| {\bf F}_{\Delta} |}{ d \Delta^{*}_{f,m} }
$ must be independent of $m$.
This imposes a self-consistency relation for the order parameters
and determines the possible superfluid phases.
Before we investigate concrete examples, we will first investigate the 
connection of heteronuclear fermionic superfluids
with spinor BECs in the next 
section.

\subsection{Ginzburg-Landau theory}\label{secglth}

In this subsection we show
 that a heteronuclear fermionic superfluid with
 Cooper pairs having a nonzero spin can be 
maped onto a spinor BEC.
For this purpose
we formulate the Ginzburg-Landau theory for our system.
We make use of the functional-integral formalism and
 as our starting point we take the action for
the Hamiltonian $\hat{H}_{1B} + \hat{V}$
 discussed in Sec. \ref{sectheory}.
The partition function $Z$ is expressed in terms of functional 
integrals over the fermionic fields $\phi_{\sigma}$ and $\chi_{\sigma}$ as
\begin{eqnarray}
Z = \int d[\phi] d[\phi^{*}] d[\chi] d[\chi^{*}]  e^{- S/\hbar},
\end{eqnarray}
where the action is given by
\begin{widetext}
\begin{eqnarray}
S[\phi^*,\phi,\chi^{*},\chi] &=& \int_{0}^{\hbar \beta} d \tau \int d\bm{x} 
 \sum_{\sigma}
{\phi}^{*}_{\sigma} (\bm{x},\tau)
 \left( i \hbar \frac{\partial}{\partial {\tau}}
- \frac{\hbar^{2} 
%\boldsymbol
{\nabla}^{2}}{2 M_{\phi}} 
+ V_{\phi,\sigma}^{\rm ex} (\bm{x}) - \mu_{\phi}
\right) 
%\delta_{m,n} +
%p
%{^{j}}F^{{z}}_{m,n}
{\phi}^{\phantom *}_{\sigma} (\bm{x},\tau)
\nonumber \\ 
&+&
\int_{0}^{\hbar \beta} d\tau \int d\bm{x} \sum_{\sigma}
{\chi}^{*}_{\sigma} (\bm{x},\tau)
\left( i \hbar \frac{\partial}{\partial {\tau}}
- \frac{\hbar^{2} 
%\boldsymbol
{\nabla}^{2}}{2 M_{\chi}} 
+ V_{\chi,\sigma}^{\rm ex} (\bm{x}) - \mu_{\chi}
\right) 
%\delta_{\sigma,\tau} +
%p
%{^{j'}}F^{{z}}_{\sigma,\tau}
{\chi}^{\phantom *}_{\sigma} (\bm{x},\tau)
\nonumber \\ 
&+&
\frac{1}{2} \int_{0}^{\hbar \beta} d\tau  \int d\bm{x}  \sum_{m} ~ 
 g_{f}  {A}^{*}_{f,m} (\bm{x},\tau)  
{A}^{\phantom *}_{f,m} (\bm{x},\tau).
\end{eqnarray}
\end{widetext}
The first two lines on the right-hand side describe free propagations of 
the fermion fields and the last line describes the two-body interaction which 
involves pairing amplitudes
\begin{eqnarray}
&& {A}_{f,m} (\bm{x},\tau) = \nonumber \\ &&
\sum_{\sigma,\sigma'} \langle f,m | f_{\phi}, \sigma; \frac{1}{2}, \sigma' \rangle
{\phi}_{\sigma} (\bm{x},\tau){\chi}_{\sigma'} (\bm{x},\tau). \nonumber \\ 
\end{eqnarray}
We consider the homogeneous case here
by setting the external potentials 
$V_{\chi,\sigma}^{\rm ex} (\bm{x})$ and $V_{\phi,\sigma}^{\rm ex} (\bm{x})$
to be zero.
We decouple the interaction by means of 
a Hubbard-Stratonovich transformation, i.e., 
we introduce a complex auxiliary field $\Delta_{f,m}(\bm{x},\tau)$ 
that couples to 
the  product of fields 
$ \sum_{\sigma,\sigma'} \langle f, m | f_{\phi},\sigma ; 
\frac{1}{2},\sigma' \rangle
\phi_{\sigma}(\bm{x},\tau) \chi_{\sigma'}(\bm{x},\tau)$.
The fourth-order term generated by the
 Hubbard-Stratonovich transformation exactly cancels 
the interaction 
term in the action  and the
resulting action $S[\Delta^*,\Delta,\phi^*,\phi,\chi^{*},\chi]$
depends only quadratically on the 
fermion fields $\phi_{\sigma} (\bm{x},\tau)$
and $\chi_{\sigma} (\bm{x},\tau)$:
\begin{eqnarray}\label{eqac24}
&&S[\Delta^{*},\Delta, \phi^{*},\phi,\chi^{*},\chi] =
- \int_{0}^{\hbar} d\tau
\int d\bm{x} \sum_{m} \frac{|\Delta_{f,m}|^{2} }{g_{f}}
\nonumber \\ &&
- \hbar 
\int_{0}^{\hbar} d\tau
\int d\bm{x}
\int_{0}^{\hbar} d\tau'
\int d\bm{x}'
\boldsymbol{\Psi}^{*} \cdot \boldsymbol{G}^{-1} \cdot 
\boldsymbol{\Psi},
\nonumber \\
\end{eqnarray}
where $\boldsymbol{\Psi}^{*}$ stands for a set of fields
\begin{eqnarray}
\boldsymbol{ {\Psi}}^{*} (\bm{x}, \tau)
=\left( 
\begin{matrix}
 {\phi}^{*}_{f_{\phi}} (\bm{x}, \tau), \ldots 
 {\phi}^{*}_{-f_{\phi}} (\bm{x}, \tau),
 {\chi}^{\phantom *}_{\frac{1}{2}} (\bm{x}, \tau),
 {\chi}^{\phantom *}_{-\frac{1}{2}} (\bm{x}, \tau)
\end{matrix}
\right), \nonumber \\
\end{eqnarray}
and properties of the system are encapsulated in the Green's function 
matrix $\boldsymbol{G}$ which is expressed in terms 
of the noninteracting Green's function $\boldsymbol{G}_{0}$ 
and self-energy $\boldsymbol{\Sigma}$ as 
\begin{eqnarray}
\boldsymbol{G}^{-1} (\bm{x},\tau; \bm{x}',\tau') =
\boldsymbol{G}_{0}^{-1} (\bm{x},\tau; \bm{x}',\tau') 
- \boldsymbol{\Sigma}.
\end{eqnarray}
The noninteracting Green's function ${\bf G}_0$ is the
diagonal $(2f_{\phi}+3)\times (2f_{\phi}+3)$ matrix
\begin{eqnarray}\label{nonint2jp3}
&&{\bf G}_0^{-1}(\bm{x},\tau;\bm{x}',\tau') = \nonumber \\ &&
\left[
\begin{matrix}
G_{\phi, 0}^{-1}(\bm{x},\tau;\bm{x}',\tau') 
{\bf 1} & 0 \\
 0 & -G_{\chi,0}^{-1}(\bm{x}',\tau';\bm{x},\tau) {\bf 1}
\end{matrix}
\right], \nonumber \\
\end{eqnarray}
where $ -G_{\alpha,0}^{-1}(\bm{x}',\tau';\bm{x},\tau) {\bf 1}$
($\alpha = \phi,\chi$)
is a $(2 f_{\phi} + 1) \times (2 f_{\phi} + 1)$ matrix
that satisfies
\begin{eqnarray}
&& G_{\alpha,0}^{-1}(\bm{x},\tau;\bm{x}',\tau') \nonumber \\ &&
= - \frac{1}{\hbar} \left\{ \hbar\frac{\partial}{\partial\tau}
    - \frac{\hbar^2 \mbox{\boldmath $\nabla$}^2}{2 M_{\alpha}}
     - \mu_{\alpha}  \right\}
 \delta(\bm{x}-\bm{x}') \delta(\tau-\tau')~.~~ \nonumber \\
\end{eqnarray}
The self-energy $\boldsymbol{\Sigma}$ is 
a  $(2f_{\phi}+3)\times (2f_{\phi}+3)$ matrix
\begin{eqnarray}\label{defthissigma}
\hbar \boldsymbol{\Sigma} 
&=&
\left[
\begin{matrix}
 0 & V \\ V^{\dagger} & 0 \\
%0 & \ldots & 0 & V_{f_{\phi},1/2} & V_{f_{\phi},-1/2} \\
%0 & \ldots & 0 & V_{f_{\phi}-1,1/2} & V_{f_{\phi}-1,-1/2} \\
% & & \vdots & & \\
%0 & \ldots & 0 & V_{-f_{\phi},1/2} & V_{-f_{\phi},-1/2} \\
%V^{*}_{f_{\phi},1/2} & \ldots & V^{*}_{-f_{\phi},1/2} & 0 & 0 \\
%V^{*}_{f_{\phi},-1/2} & \ldots & V^{*}_{-f_{\phi},-1/2} & 0 & 0 \\
\end{matrix}
\right], \nonumber \\
\end{eqnarray}
where $V$ is the
$(2 f_{\phi} + 1) \times 2$ matrix 
 given by
$
V_{\sigma, \sigma'} = 
\sum_{m}
 \langle f,m | f_{\phi}, \sigma; 1/2, \sigma' \rangle \Delta_{f,m}
$. 
Since the action (\ref{eqac24}) depends only quadratically on 
the fermion fields, we can integrate them out and
 obtain an effective action
\begin{eqnarray}\label{effectivea}
S^{\rm eff}[\Delta^*,\Delta] &=&
        - \int_0^{\hbar\beta} d\tau \int d\bm{x}~
\sum_{m}
            \frac{|\Delta_{f,m}(\bm{x},\tau)|^2}{g_{f}}
\nonumber \\ &&    - \hbar {\rm Tr}[\log (-{\bf G}^{-1})]~.
\end{eqnarray}
The last term in Eq. (\ref{effectivea})
 can be expanded in powers of $\Delta$ by using
\begin{eqnarray*}
{\bf G}^{-1}={\bf G}_0^{-1} - \mbox{\boldmath $\Sigma$}
     = {\bf G}_0^{-1}(1 - {\bf G}_0 \mbox{\boldmath $\Sigma$}),
\end{eqnarray*}
where the self-energy $\hbar \boldsymbol{\Sigma}$ is given by
Eq. (\ref{defthissigma}),
and therefore
\begin{eqnarray}\label{eq22}
- \hbar {\rm Tr}[\log (-{\bf G}^{-1})] &=&
 - \hbar {\rm Tr}[\log (-{\bf G}_0^{-1})]
\nonumber \\ &&  + \hbar \sum_{m=1}^{\infty} \frac{1}{m}
                          {\rm Tr}[({\bf G}_0 \mbox{\boldmath $\Sigma$})^m]~.
\end{eqnarray}
The second-order $m=2$ term in the last term on the right-hand-side of 
Eq. (\ref{eq22}) is given by
\begin{widetext}
\begin{eqnarray}\label{kinkinkin}
&& \frac{\hbar}{2} {\rm Tr}\left[
\left( {\bf G}_{0}  \mbox{\boldmath $\Sigma$} \right)^{2}
\right]
=  \frac{\hbar}{2}
 \int d{\tau} \int d\bm{x}
\int d{\tau}' \int d\bm{x}'
\int d{\tau}'' \int d\bm{x}''
\int d{\tau}''' \int d\bm{x}'''
\nonumber \\ &&
{\rm tr} \left[
 {\bf G}_{0} (\bm{x},\tau;\bm{x}',\tau')
\mbox{\boldmath $\Sigma$} (\bm{x}',\tau';\bm{x}'',\tau'')
 {\bf G}_{0}   (\bm{x}'',\tau'';\bm{x}''',\tau''')
\mbox{\boldmath $\Sigma$}  (\bm{x}''',\tau''';\bm{x},\tau)
\right], \nonumber \\
\end{eqnarray}
\end{widetext}
where the trace operation ${\rm tr}[\ldots]$
on the  right-hand side 
 means that we  only  take
the sum of the diagonal elements of the $(2f_{\phi}+3)\times(2f_{\phi}+3)$ matrix
${\bf G}_{0} \boldsymbol{\Sigma}{\bf G}_{0} \boldsymbol{\Sigma}$.
In the following we assume $\Delta_{f,m}$ to be 
independent of space and (imaginary) time.  This allows us to separate 
the imaginary-time and space 
integrations from the matrix trace.
For equal chemical potentials and equal masses 
we have $G_{\phi,0}^{-1}(\bm{x},\tau;\bm{x}',\tau') = G_{\chi,0}^{-1}(\bm{x},\tau;\bm{x}',\tau')$, and
 the integrals over the Green's functions reduce to those of 
the standard BCS theory. 

The new contributions come from the matrix trace.
For the second-order term we find after some straightforward algebraic
manipulations that
\begin{eqnarray}
{\rm tr} \left[ V^{\dagger} V \right] =
{\rm tr} \left[ V V^{\dagger} \right] =
 \sum_{m} |\Delta_{f,m}|^{2}.
\end{eqnarray}
To evaluate the fourth-order term, we make use of
\begin{eqnarray}
{\rm tr} \left[ V^{\dagger} V  V^{\dagger} V \right] &=& 
{\rm tr} \left[ V V^{\dagger} V  V^{\dagger} \right] \nonumber \\
&=&
\left( {\rm tr} \left[ V^{\dagger} V  \right] \right)^{2}
- 2 {\rm Det}  \left[ V^{\dagger} V  \right]
\nonumber \\
&=&
\frac{1}{2} \left[ \left( \sum_{m} |\Delta_{f,m}|^{2} \right)^{2}
+ \frac{{\bf F}_{\Delta} \cdot{\bf F}_{\Delta} }{{f^{+}}^{2}}
\right]. \nonumber \\
\end{eqnarray}
Combining these with conventional BCS theory we 
 find that to the fourth order the thermodynamic potential,
for equal masses and chemical potentials, is given by
\begin{eqnarray}\label{GLthpot}
\Omega = 
\alpha_{f}(T) \sum_{m} |\Delta_{f,m}|^{2}
+ \beta \left(  \left(\sum_{m} |\Delta_{f,m}|^{2} \right)^{2}
+  \frac{|{\bm F}_{\Delta}|^{2}}{{f^{+}}^{2}}
\right), \nonumber \\
\end{eqnarray}
where 
$
\alpha_{f}(T) = 
N(0) \log{T^{f}_{\rm c}/T}$,
$\beta = 7 \zeta(3) N(0)/ 16 (\pi k_{\rm B} T^{f}_{\rm c})^{2} $, 
$N(0) = 2 M k_{\rm F} / (2 \pi \hbar)^{2} $ 
is the density of states at the Fermi energy,
which is the same for both species since $M=M_{\phi} = M_{\chi}$ \cite{Melo}.
The critical temperature is given by
\begin{eqnarray}
T^{f}_{\rm c} = \frac{8 \epsilon_{\rm F}}{\pi k_{\rm B}} e^{\gamma - 2} 
e^{-\pi/2 |k_{\rm F} a_{f}|}.
\end{eqnarray}
The form of the thermodynamic potential in Eq. (\ref{GLthpot})
is essentially the same as that 
of the homogeneous spinor BECs at zero magnetic field. 
Here, however, since $\beta > 0$ the ground state is
 unmagnetized, i.e., ${\bm F}_{\Delta} = 0$. Physically this
result arises from the Pauli principle which forbids the 
fermionic constituents of the composite bosons to occupy the same state.

Moreover, we only find two kinds of fourth-order terms, i.e., the 
particle density and spin density, regardless of the 
value of the spin $f$. In contrast, for the case of spinor BECs
there are $f+1$ different fourth-order terms.
Therefore, all unmagnetized phases predicted in spin-$f$ BECs
 are degenerate in the present case.

\section{Results}\label{secres}
As shown above, the heteronuclear fermionic superfluid has a Ginzburg-Landau
free energy which is similar in form to that of a spinor BEC. We 
may thus expect  similarities between the two systems
insofar as mean-field theory is valid.
To investigate this problem, we solve the gap equation (\ref{gapeqn}) for 
the case of two spin-$1/2$ species whose total spin is either 
zero or one with the corresponding $s$-wave scattering length being 
$a_{0}$ or $a_{1}$.
The phase which the gas condenses into 
is determined by the  $s$-wave scattering length
with the largest amplitude.
For $f=0$ we immediately conclude that $ {\bf F}_{\Delta} = 0$.
The Bogoliubov spectrum is the same as that of the conventional BCS theory.
For total angular momentum $f=1$ we have
\begin{eqnarray}
 |{\bf F}_{\Delta}|^{2}
&=& \left( \sum_{m} |\Delta_{1,m}|^{2} \right)^{2}
- 3 |\Theta_{1} |^{2}
,
\end{eqnarray}
where  the scalar quantity $\Theta_{1} = 
(-\Delta_{1,0}^{2} + 2 \Delta_{1,1} \Delta_{1,-1} )/\sqrt{3}$
is the same in form as the spin-singlet state pair amplitude of a
spin-$1$ BEC \cite{Ueda2000}.
If all $\Delta_{1,m}$ are nonzero,
 we find from the gap equation that either 
$\Theta_{1} = 0$ or 
\begin{eqnarray}
\frac{1}{\Delta_{1,m}} \frac{\partial \Theta_{1} }{\partial \Delta^{*}_{1,m} } =
\frac{1}{\Delta_{1,m'}} \frac{\partial \Theta_{1} }{\partial \Delta^{*}_{1,m'} }
\end{eqnarray}
holds for $m,m' = 0,\pm 1$.
It follows 
from the latter condition 
that $|\Delta_{1,1}| = |\Delta_{1,-1}|$. Moreover, if we represent the
order parameters as $\Delta_{1,m} = |\Delta_{1,m}| e^{i \theta_{1,m}}$,
we find from the 
gap equation that $2 \theta_{1,0} = \theta_{1,1} + \theta_{1,-1} + \pi$.
We also define the angle $\beta$ from the relation $\Delta_{1,0} = 
(|\Delta|/\sqrt{2}) \cos{\beta}$, where
$|\Delta|^{2} =\sum_{m} |\Delta_{f,m}|^{2}$.
Combining these, we find that 
\begin{eqnarray}
&&\left(
\Delta_{1,1} ,
\Delta_{1,0} ,
\Delta_{1,-1} 
\right) = \nonumber \\ &&
 |\Delta|e^{i \theta_{1,0}} 
\left(
\frac{1}{\sqrt{2}} e^{i \alpha } \sin{\beta}, 
\cos{\beta}, 
-\frac{1}{\sqrt{2}} e^{-i \alpha} \sin{\beta} 
\right), \nonumber \\
\end{eqnarray}
where $\alpha = \theta_{1,1} - \theta_{1,0}$.
This order parameter also describes the polar phase
of the spin-$1$ Bose gas.
From the condition $\Theta_{1} = 0$, we obtain 
the phase relation $2 \theta_{1,0} = \theta_{1,1} + \theta_{1,-1}$
and the amplitude relation $|\Delta_{1,0}|^{2} = 
2 |\Delta_{1,1}| |\Delta_{1,-1}|$. From these results
we find that the ferromagnetic order parameter
is given by
\begin{eqnarray}\label{eq41}
&& (\Delta_{1,1},\Delta_{1,0}, \Delta_{1,-1}) =  \nonumber \\ &&
|\Delta| e^{i \theta_{1,0}} 
\left(
e^{ i \alpha} \cos^{2}{\frac{\beta}{2}},
\sqrt{2} \sin{\frac{\beta}{2}} \cos{\frac{\beta}{2}},
e^{-i \alpha} \sin^{2}{\frac{\beta}{2}}
\right). \nonumber \\
\end{eqnarray}
It can be shown that Eq. (\ref{eq41}) is indeed a solution to
the gap equation (\ref{gapeqn}). 
Comparing the free energies
of the polar and ferromagnetic phases we
conclude that the ground state is polar.
Because the order parameter  of the polar phase has a ${\bm Z}_{2}$ symmetry,
i.e., is invariant under the  transformations
$\alpha \rightarrow \alpha + \pi$ and $\beta \rightarrow \pi - \beta $,
the mean-field theory predicts a quantum phase transition 
when we change the scattering lengths 
from  $a_{1} > a_{0}$ to $a_{1} < a_{0}$, and vice versa. 

To find solutions of the gap equation
for higher values of the spin $f_{\phi} > 1/2$
it is convenient to make use of the one-to-one 
correspondence between the fermionic heteronuclear gas 
and the spinor BEC derived in the previous section.
The solutions  for total spin $2$ and spin $3$ are of special 
interest because the correspond to 
 the experimentally realized case 
of a mixture of \mbox{$^{171}$Yb} (spin $1/2$) and 
\mbox{$^{173}$Yb} (spin $5/2$).
The spinor BECs with spin $2$ and $3$ have been
described in the literature
\cite{Hospin2bec, Uedaspin2bec,Santosspin3bec,Hospin3bec}.
and it is straightforward to verify that the possible ground-state
solutions of the spin $2$ and spin  $3$ BECs 
also satisfy the gap equation (\ref{gapeqn}).
For example, for the case of the spin $2$ BEC
 cyclic and
 polar phases are possible \cite{Hospin2bec, Uedaspin2bec}.
These phases can be distinguished  from each other by the 
value of 
$\Theta_{2} = 
(\Delta_{2,0}^{2} - 2 \Delta_{2,1} \Delta_{2,-1} + 2 \Delta_{2,-2} \Delta_{2,2})/\sqrt{5}
$, which describes the formation of singlet pairs of 
spin-$2$ atoms. 
For spin $3$ the solutions are more complex and have been 
discussed by  Santos and Pfau \cite{Santosspin3bec}
and Diener and Ho \cite{Hospin3bec}. Also a complete classification 
of states and vortex excitations 
have been discussed in Refs. \cite{Yipclass,Barnett}.

\section{Possible experimental realizations}\label{secexp}
The critical temperature needed to reach the weak-coupling limit is beyond 
experimental reach and in order to increase the critical temperature to 
experimentally accessible values we need to make use of optical Feshbach resonances
\cite{Fedichev}
so that the effective interactions between the atoms are enhanced. 
Currently, optical Feshbach resonances
 for alkali atoms  \cite{Fatemi, Theis}
 are not as effective as magnetic Feshbach resonances 
due to large atomic losses. It has been predicted
 \cite{Julienne0,Julienne1}, however, that for 
ground-state alkaline-earth-metal atoms optical Feshbach 
resonances can be used to tune the scattering length over a wide 
range of values without suffering from the large atomic losses.
As mentioned in the introduction, the fermionic isotopes
of Yb have already been trapped optically
\cite{Takahashi1}, which makes them a prime candidate for 
realizing a  heteronuclear fermionic superfluid with 
spin degrees of freedom. 
At present, experimental efforts are underway to find appropriate 
bound states in the long-lived  ($>10$sec)
excited state potentials 
$^{1}S_{0} + {^{3}}P_{i=0,2}$ of Yb that can be used for 
optical Feshbach resonances \cite{TakahashiPC}.
If these efforts are successful we expect that the ground-states
of the \mbox{$^{171}$Yb}  and \mbox{$^{173}$Yb} mixture are
the same as the unmagnetized phases of the
 spin $2$ and $3$ spinor BECs. 
Moreover, in the absence of symmetry breaking 
perturbations, the excitation spectrum for 
the unmagnetized 
ground-states is given by Eqs. (\ref{bogdisp}) and  (\ref{bogdisp2})
with $|{\bm F}_{\Delta}|=0$
and has the same form as the standard BCS theory.
For a realistic gas, however, 
symmetry-breaking terms such as 
magnetic fields and dipole-dipole or 
intraspecies interactions are expected to
lift the degeneracy and we plan to investigate this in future research.

\section{Summary and outlook}\label{secsum}
In summary, we have presented a theory of superfluidity in 
heteronuclear Fermi-Fermi mixtures. 
We have shown that
the heteronuclear fermionic superfluid has a Ginzburg-Landau
free energy which is similar in form to that of a spinor BEC
and, therefore,  analogous mean-field ground states are expected
for these two systems. This raises an  important question concerning 
the nature of the many-body ground state in our case.
For the bosonic case it has been shown that the exact many-body state is,
in fact, not the polar state but a condensate
of spin singlet pairs \cite{Law,koashi,Yipyip}. 
Explicitly, in the absence of a magnetic field
 the exact bosonic many-body state can be written as
$|\psi\rangle \propto
\left[
((\hat{a}_{0}^{\dagger})^2 - 2 \hat{a}_{1}^{\dagger}\hat{a}_{-1}^{\dagger})/\sqrt{3}
\right]^{N/2} |0 \rangle
$, where 
$a_{m}^{\dagger}$ creates a boson in 
hyperfine state $m=0,\pm 1$ with zero linear momentum  and 
$N$ is the number of bosons.
To perform a similar many-body analysis for
the heteronuclear Fermi superfluids is beyond the scope of the
present paper and 
will be left for future research. It is clear, however, that
if the similarity at the mean-field level extends to the many-body case, 
the exact ground state is expected to be a condensate of spin-singlet 
states involving a quartet of fermions. 
Quartet superfluidity is of interest 
in  strongly interacting quantum liquids \cite{Roepke,Kamei} and  
in the formation of two-pion states \cite{alm}. In these works, however,
only single species gases were considered for which a fourfold degeneracy
and moderate coupling are indispensible. Our work 
demonstrates the possibillity of creating a quartet condensate using only 
spin-$1/2$ particles because  the 
Pauli principle does not prevent different species from occupying
the same spin state.  This unique feature of heteronuclear fermionic superfluidity merits further theoretical and experimental investigations.

\section{Acknowledgements}
This work was supported by a Grant-in-Aid for Scientific Research 
(Grant No. 17071005) and by a 21st Century COE program on 
``Nanometer-Scale Quantum Physics''
      from the Ministry of Education, Culture,
Sports, Science, and Technology of Japan. 
D.D. acknowledges  H. Takahashi and his group for stimulating 
discussions concerning the experimental implementations 
of this work and support by the Japan Society for the 
Promotion of Science (Project No. 18.06716).

\appendix
\section{The Bogoliubov spectrum of a heteronuclear fermionic superfluid}\label{appBD}
Substituting the mean-field expression in Eq.
(\ref{mfV}) into Eq. (\ref{eq1}),
we obtain the following Hamiltonian for a homogeneous 
system:
\begin{eqnarray}
\hat{H} &=& 
\sum_{\bf k}
\left(
\begin{matrix}
\phi^{\dagger}_{f_{\phi}} \\ 
\vdots \\
\phi^{\dagger}_{-f_{\phi}} \\
\chi_{1/2} \\
\chi_{-1/2}
\end{matrix}
\right)
\left(
\begin{matrix}
(\xi_{\phi,{\bf k}} - \lambda ) {\bf 1} & V \\
V^{\dagger} & (- \xi_{\chi,{\bf k}} - \lambda ) {\bf 1}
\end{matrix}
\right)
\left(
\begin{matrix}
\phi^{*}_{f_{\phi}} \\ 
\vdots \\
\phi^{*}_{-f_{\phi}} \\
\chi^{\dagger}_{1/2} \\
\chi^{\dagger}_{-1/2}
\end{matrix}
\right)
\nonumber \\ 
&&+ \sum_{\bf k} 2 \xi_{\chi,\bf k}
- \sum_{m} \frac{|\Delta_{f,m}|}{g_{f}}, \nonumber \\
\end{eqnarray}
where we define the $(2 f_{\phi} + 1)\times 2$ matrix
\begin{eqnarray}\label{defA2}
V = -
\left( \begin{matrix}
 V_{j,1/2} & V_{j,-1/2} \\
 V_{j-1,1/2} & V_{j-1,-1/2} \\
 \vdots  & \vdots   \\
 V_{-j,1/2} & V_{-j,-1/2} \\
\end{matrix} \right), \nonumber \\
\end{eqnarray}
with matrix elements
$V_{\sigma, \sigma'} = 
\sum_{m}
 \langle f,m | j, \sigma; 1/2, \sigma' \rangle \Delta_{f,m}$.
The eigenvalue problem associated with this Hamiltonian can thus 
be written in a compact form as
\begin{eqnarray}\label{eveqn}
\left(
\begin{matrix}
(\xi_{{\bf k}} - \lambda' ) {\bf 1} & V \\
V^{\dagger} & (- \xi_{{\bf k}} - \lambda' ) {\bf 1}
\end{matrix}
\right) \left( \begin{matrix}
u \\ v
\end{matrix} \right) = 0. \nonumber \\
\end{eqnarray}
Here, we have introduced the variable 
$\xi_{\bf k} = (\xi_{\phi,{\bf k}} + \xi_{\chi,{\bf k}})/2$
and the shifted eigenvalue
$\lambda' = \lambda - (\xi_{\phi,{\bf k}} - \xi_{\chi,{\bf k}})/2$.
The vector $u$ has $2j+1$ components
and $v$ has two components.
The highest order of $\lambda'$ in the eigenvalue eqation is 
$\lambda'^{2j + 3}$, which determines the
maximum number of modes in the system.
After working out the multiplications,
we obtain the following coupled equations
\begin{eqnarray}\label{eqa4}
(\xi_{\bf k} - \lambda') {\bf 1} u + V\cdot v  = 0 
\end{eqnarray}
and
\begin{eqnarray}\label{eqa5}
(- \xi_{\bf k} - \lambda') {\bf 1} v + V^{\dagger}\cdot u  = 0 
\end{eqnarray}
We solve Eq. (\ref{eqa4}) formally for 
$u$ and substitute the solution into
Eq. (\ref{eqa5}) to obtain the following 
equation for $v$:
\begin{eqnarray}
\left\{
(- \xi_{\bf k} - \lambda') {\bf 1}
+  V^{\dagger}\cdot \frac{1}{ \xi_{\bf k} - \lambda'} {\bf 1}
\cdot V
\right\} v = 0.
\end{eqnarray}
The eigenvalues of this equation can be obtained by 
solving the characteristic equation,
\begin{eqnarray}\label{chareqn}
{\rm Det}
\left[ (\xi_{\bf k} - \lambda')(- \xi_{\bf k} - \lambda') {\bf 1}
+ V^{\dagger}V \right] = 0. \nonumber \\
\end{eqnarray}
Similarly, we can obtain for $u$ that 
\begin{eqnarray}
\left\{
(\xi_{\bf k} - \lambda') {\bf 1}
+  V \cdot \frac{1}{ - \xi_{\bf k} - \lambda'} {\bf 1}
\cdot V^{\dagger}
\right\} u = 0. \nonumber \\
\end{eqnarray}
The eigenvalues for this can be obtained by 
solving the characteristic equation 
\begin{eqnarray}\label{eqa10}
{\rm Det}
\left[ (\xi_{\bf k} - \lambda')(- \xi_{\bf k} - \lambda') {\bf 1}
+ V V^{\dagger} \right] = 0. \nonumber \\
\end{eqnarray}
While Eq. (\ref{chareqn}) is a $2 \times 2$ eigenvalue equation,
Eq. (\ref{eqa10}) is a $(2 f_{\phi} + 1) \times (2 f_{\phi} + 1)$ eigenvalue
equation which has $2 f_{\phi} - 1$ degenerate solutions
$\lambda = \pm \xi_{\bf k}$ in addition to the nontrivial solutions.
 The nontrivial solutions follow from Eq. (\ref{chareqn}). We have,
\begin{eqnarray}\label{vdagv}
V^{\dagger} V = 
\left(
\begin{matrix}
\sum_{j'} V_{j',1/2}^{*} V_{j',1/2} &
\sum_{j'} V_{j',1/2}^{*} V_{j',-1/2} \\
\sum_{j'} V_{j',-1/2}^{*} V_{j',1/2} &
\sum_{j'} V_{j',-1/2}^{*} V_{j',-1/2}
\end{matrix}
\right) \nonumber \\
\end{eqnarray}
Using this, we obtain the eigenvalues $\lambda$ from Eq.
(\ref{chareqn}),
\begin{eqnarray}
&&\lambda = \frac{\xi_{\phi, {\bf k}} - \xi_{\chi, {\bf k}}}{2}  
\nonumber \\ &&
 \pm \sqrt{
\left( \frac{\xi_{{\bf k}}}{2} \right)^{2}
+ \frac{1}{2} {\rm tr} [V^{\dagger} V]
\pm \frac{1}{2}
\sqrt{
\left( {\rm tr}[V^{\dagger} V] \right)^{2}
- 4 {\rm Det}[V^{\dagger} V]
}
}. \nonumber \\
\end{eqnarray}
Next, we express the trace and determinant terms appearing in the above 
eigenvalues in terms of physical quantities. Firstly, 
combining Eq. (\ref{vdagv}) with Eq. (\ref{defA2}), we obtain
\begin{eqnarray}
{\rm tr}\left[ V^{\dagger} V \right] = 
\sum_{j',\sigma'} |V_{j',\sigma'}|^{2}
= \sum_{m} |\Delta_{f,m}|^{2}
\end{eqnarray}
The other term is given by 
\begin{eqnarray}
&& \left( {\rm tr}\left[ V^{\dagger} V \right] \right)^{2} - 
4 {\rm Det}[V^{\dagger} V] =
 \frac{|{\bm F}_{\Delta}|^{2}}{{f^{+}}^{2}}
, \nonumber \\
\end{eqnarray}
where ${\bm F}_{\Delta}$ is given by Eq. (\ref{cpspin}).

\bibliographystyle{apsrev}

\end{document}